\documentclass[letterpaper, 10 pt, conference]{ieeeconf}  

\IEEEoverridecommandlockouts                              

\usepackage{graphics} 
\usepackage{epsfig} 
\usepackage{mathptmx} 
\usepackage{times} 
\usepackage{amsmath} 
\usepackage{amssymb}  
\usepackage{algorithm}
\usepackage{epstopdf}
\usepackage{cite}
\usepackage{color}
\usepackage[noend]{algpseudocode}
\usepackage{bm}
\usepackage{comment} 
\usepackage[english]{babel}
\usepackage[latin1]{inputenc}
\usepackage{times}
\usepackage[T1]{fontenc}
\usepackage{booktabs}
\usepackage{amsmath}
\usepackage{graphicx}
\usepackage{epstopdf}
\usepackage{float}
\usepackage{enumerate}
\usepackage{subfigure}
\usepackage{color}
\usepackage{cite}
\usepackage{setspace}
\usepackage{mathptmx}
\usepackage{times} 
\usepackage{amsmath} 
\usepackage{amssymb}  
\usepackage{algorithm}
\usepackage{bm}
\usepackage[usestackEOL]{stackengine}
\usepackage{epstopdf}
\usepackage{nccmath}
\usepackage{tabularx,ragged2e,booktabs}
\usepackage[mathscr]{eucal}
\newcolumntype{L}{>{\RaggedRight\arraybackslash}X} 

\usepackage{array}
\newcolumntype{L}[1]{>{\raggedright\let\newline\\\arraybackslash\hspace{0pt}}m{#1}}
\newcolumntype{C}[1]{>{\centering\let\newline\\\arraybackslash\hspace{0pt}}m{#1}}
\newcolumntype{R}[1]{>{\raggedleft\let\newline\\\arraybackslash\hspace{0pt}}m{#1}}

\title{\LARGE \bf
Sparse Wide-Area Control of Power Systems using Data-driven Reinforcement Learning 
}

\author{Amirhassan Fallah Dizche, Aranya Chakrabortty, and  Alexandra Duel-Hallen
\thanks{This research is partly supported by the US National Science Foundation under grant ECCS 1544871.}
\thanks{Amirhassan Fallah Dizche, Aranya Chakrabortty, and  Alexandra Duel-Hallen are are with the Department of Electrical Engineering, North Carolina State University, Raleigh,
NC, USA 
        {\tt\small afallah@ncsu.edu, aranya.chakrabortty@ncsu.edu, sasha@ncsu.edu }}%
}

\begin{document}

\maketitle
\thispagestyle{empty}
\pagestyle{empty}

\begin{abstract}

In this paper we present an online wide-area oscillation damping control (WAC) design for uncertain models of power systems using ideas from reinforcement learning. We assume that the exact small-signal model of the power system at the onset of a contingency is not known to the operator and use the nominal model and online measurements of the generator states and control inputs to rapidly converge to a state-feedback controller that minimizes a given quadratic energy cost.  However, unlike conventional linear quadratic regulators (LQR), we intend our controller to be sparse, so its implementation reduces the communication costs. We, therefore, employ the gradient support pursuit (GraSP) optimization algorithm to impose sparsity constraints on the control gain matrix during learning. The sparse controller is thereafter implemented using distributed communication. Using the IEEE 39-bus power system model with 1149 unknown parameters, it is demonstrated that the proposed learning method provides reliable LQR performance while the controller matched to the nominal model becomes unstable for severely uncertain systems.   

\end{abstract}

\section{INTRODUCTION}

Over the past few years, the occurrence of a series of blackouts in different parts of the world has led power system utility owners to look beyond the traditional approach of controlling the grid via local feedback and instead transition to system-wide control, often referred to as wide-area control (WAC). Several papers on WAC design for damping of electromechanical oscillations have been reported in the recent literature \cite{1,2,3,4 }.The basic  approach is to linearize the system model around a given operating point, and design linear state-feedback or output-feedback LQR controllers for taking damping control action via the generator excitation control system. However, designing a traditional LQR controller is not suitable for WAC since it demands a dense all-to-all communication graph between every pair of generators. To save on communication costs, sparse control designs for WAC have been reported in several recent papers such as \cite{5}, \cite{jsac}, \cite{abhi}. But a common limitation among all these designs is that they are based on perfect knowledge of the grid model. 

 In reality, however, the operating point of a grid may move over wide ranges, and therefore using just one fixed controller might not be optimal. The problem is becoming more notable with increasing penetration of renewables, power electronics, and active loads, whose dynamic characteristics vary constantly over time. One way to counteract these uncertainties would be to design a robust WAC. The challenge, however, is that with millions of electric vehicles and inverter-based generation points being envisioned to be integrated to the US grid in very near future, primarily in a completely plug-and-play fashion, it is extremely difficult to quantify a reliable upper bound for these uncertainties that can be used for robust control designs. Recent papers such as \cite{ACC5, ACC8, feieracc} have proposed robust sparse control, but those designs usually work for fairly limited amount of uncertainty. Operators are, therefore, more interested in learning the power system model using online measurements available from sophisticated sensors such as Phasor Measurement Units (PMU) after contingencies, and in developing real-time control actions that result from learning. 

Motivated by this problem, in this paper we present a LQR-based WAC design using online reinforcement learning (RL). RL has been shown to be an excellent candidate for online optimal control design under model uncertainty in several recent papers such as \cite{RL2,RL4,RL6,OP1}. Other variants of online learning such as adaptive dynamic programming (ADP) \cite{ADP,ADP1}, Q-learning \cite{q}, and integral concurrent learning \cite{D1}, for both continuous-time and discrete-time dynamic systems have also been proposed. In this paper, we adopt the RL design proposed in \cite{V17}, whereby online measurements of generator states and control inputs are used to learn an optimal LQR controller, given a choice of the objective function. However, the algorithm in \cite{V17} has very long convergence time due to the assumption of completely unknown system model. In this paper, we exploit the knowledge of the approximate, or nominal, model to speed up convergence significantly. Using online measurements instead of labeled data-sets categorizes the proposed algorithm as unsupervised learning while using the objective function value as reinforcement signal will make it RL.

In order to reduce the communication cost, we integrate the RL design with Gradient Support Pursuit (GraSP) that imposes sparsity constraints on the control gain matrix \cite{grasp}. The proposed algorithm incorporates the advantages of RL control and offline sparse controllers. This algorithm learns a sparse controller, thus simultaneously satisfying  the communications cost constraint and overcoming the model uncertainty. The proposed design is carried out in two sequential stages: (1) following a contingency, state estimates generated by decentralized Kalman filters at each generator, as well as the generator control inputs stream in to a central coordinator that serves as a `critic' which simultaneously learns the sparse optimal controllers $\bm{K}_{SP}$ and applies the corresponding control input $u$; (2) Once the $\bm{K}_{SP}$-learning loop converges, the controller is implemented by a distributed sparse communication topology connecting the selected sets of generators. We validate these two stages using simulations of the IEEE 39-bus 10-generator power system model with 1149 unknown parameters. We highlight the numerical trade-offs of the two stages for learning versus implementation for different levels of uncertainty. 

The \textit{main contributions} of this paper are:
\begin{itemize}
\item Reduce the convergence time of online RL control algorithm by exploiting the knowledge of the nominal model for WAC of power systems.
\item Develop a a sparsity-constrained online learning control algorithm that reduces the communication cost.
\end{itemize}

The rest of the paper is organized as follows. Section II formulates the proposed sparse WAC problem. Section III briefly reviews the use of RL for LQR designs and presents the main sparse learning algorithm by integrating RL with GraSP. Section IV presents simulation results and numerical analysis. Section V concludes the paper.

\section{Problem Statement}

\subsection{Power System Model}

Consider a power system network with $n$ synchronous machines. Each machine is considered to be modeled by its  \textit{flux-decay} model \cite{kundur}, which is a common choice for designing wide-area damping controllers using excitation control. The model for the $i^{th}$ generator can be written as
\begin{fleqn}
\begin{align}
\dot{\delta_i} = & \omega_i\\
M_i\dot{\omega_i} = & P_{mi} - d_i \omega_i - \frac{|\bm{V}_i|E_i}{x^{'}_{di}}\sin(\delta_i-\angle \bm{V}_i)  \\ \nonumber
& + \frac{|\bm{V}_i|^{2}}{2}\Big( \frac{1}{x^{'}_{di}} - \frac{1}{x_{qi}} \Big)\sin(2\delta_i-2\angle \bm{V}_i) \\
\tau_{doi} \dot{E_i} = &- \frac{x_{di}}{x^{'}_{di}}E_i + \Big(\frac{x_{di}}{x^{'}_{di}} -1  \Big)|\bm{V}_i| \cos(\delta_i-\angle \bm{V}_i) + V_{fdi},
\end{align}
followed by active and reactive power balance equations
\begin{align}
P_i  = &  \frac{E_i|\bm{V}_i|}{x^{'}_{di}} \sin(\delta_i-\angle \bm{V}_i) \\ \nonumber & -\frac{|\bm{V}_i|^{2}}{2}\Big( \frac{1}{x^{'}_{di}} - \frac{1}{x_{qi}} \Big)\sin(2\delta_i-2\angle \bm{V}_i)\\ 
Q_i = &\frac{E_i|\bm{V}_i|}{x^{'}_{di}} \cos(\delta_i-\angle \bm{V}_i) \\
&- |\bm{V}_i|^{2} \Bigg(\frac{\sin^{2}(\delta_i-\angle \bm{V}_i)}{x_{qi}}  -  \frac{\cos^{2}(\delta_i-\angle \bm{V}_i)}{x^{'}_{di}} \Bigg). \nonumber
\end{align}

\end{fleqn}

\noindent Equations (1)-(2) represent the swing dynamics, and (3) represents the electro-magnetic dynamics of the $i^{th}$ generator. $\bm{V}_i$ is the voltage phasor at the generator bus, $P_i$ and $Q_i$ are the active and reactive power outputs of the generator, $V_{fdi}$ is the exciter voltage, and the remaining constants denote various model parameters whose definitions can be found in \cite{kundur}. The generator model is coupled with the model of an exciter consisting of an automatic voltage regulator (AVR) and a power system stabilizer (PSS) whose combined dynamics can be written as

\begin{fleqn}
\begin{align}
 \tau_{e_i} \dot{V}_{fdi} = & -V_{fdi} + V^{\bigstar}_{fdi} + K_{ai}( |\bm{V}_i| - |\bm{V}_i|^{\bigstar} - \nu_i + \gamma_i )\\
 \dot{\zeta_i} = & \bm{A}_{pss}\zeta_i + \bm{B}_{pss}\omega_i,\mbox{ } \nu_i = C_{pss}\zeta_i + D_{pss}\omega_i 
\end{align}
\end{fleqn}
\noindent where superscript $^{\bigstar}$ means set-point. The signal $\gamma_i$ serves as a control input representing an additional voltage reference signal to the AVR that can be designed to add damping to the  slow  or  inter-area  oscillation  modes using state feedback from all generators spread across the grid. These controllers are referred to as wide-area controllers (WAC).

Our control design does not necessarily need the generators to follow this simple model. Detailed models of generators are allowed, provided all the generator states can be measured or estimated (a short description of decentralized state estimation will be given shortly). In general, we assume that the nonlinear model of the $i^{th}$ generator has $n_i$ states $\xi_i = [\delta_i, \omega_i,  E_i, \dots] \in \mathbb R^{n_i}$, and one scalar control input $\gamma_{i}$ as in (6), which is the field excitation voltage. Let the pre-disturbance equilibrium of the $i^{th}$ generator be $\xi_i^{\ast}=[{\delta_i}^{\ast}, {\omega_i}^{\ast}, {E_i}^{\ast}, \dots]$. The differential-algebraic model of the generators and the power flow is converted to a state-space model using Kron reduction \cite{kundur}, and linearized about $\xi_i^{\ast}, i=1, 2,\dots, n$.  The small-signal model of the system with the $i^{th}$ state defined as $\bm{x}_i=\xi_i-\xi_i^{\ast}$, is written as
\begin{multline} \label{eq7}
\begin{bmatrix}
\bm{\dot{x}_{1}}(t)\\
\bm{\dot{x}_{2}}(t)\\
\vdots\\
\bm{\dot{x}_{n}}(t)\\
\end{bmatrix} =
\begin{bmatrix}
\bm{A_{11}} & \bm{A_{12}} & \cdots &\bm{A_{1n}}\\
\bm{A_{21}} & \bm{A_{22}} & \cdots &\bm{A_{2n}}\\
\vdots\\
\bm{A_{n1}} & \bm{A_{n2}} & \cdots &\bm{A_{nn}}
\end{bmatrix}
\begin{bmatrix}
\bm{x}_{1}(t)\\
\bm{x}_{2}(t)\\
\vdots\\
\bm{x}_{n}(t)\\
\end{bmatrix} + \\
\begin{bmatrix}
\bm{B}_{1} & & &\\
  & \bm{B}_{2}& &\\
   & & \ddots &\\
    & & &\bm{B}_{n}\\
\end{bmatrix}
\begin{bmatrix}
u_{1}(t)\\
u_{2}(t)\\
\vdots\\
u_{n}(t)\\
\end{bmatrix}.
\end{multline}
\noindent
Note that the state vector $\bm{x}_i$ includes the AVR and PSS states from (6)-(7) linearized around their respective equilibria. The small-signal control input is given by $u_i=\Delta \gamma_i$. The power system model (\ref{eq7}) can be written in compact form by stacking state and input vectors as
\begin{equation} \label{eq8}
\bm{\dot{x}}(t) = \bm{Ax}(t) + \bm{Bu}(t),\quad \bm{x}(0) = \bm{x}_{0}.
\end{equation}
 Model parameters and operating conditions in the grid change frequently between contingencies, and therefore the assumption of the exact knowledge of $\bm{A}$ and $\bm{B}$ is impractical. In the current state of art, utilities use offline models that may have been constructed years ago and simply depend on the inherent robustness of the grid to save the closed-loop response even if the control input $\bm{u}$ is not properly matched with the actual $(\bm{A}, \bm{B})$ matrices that apply to that situation. With rapid increase in renewable penetration and their power electronic interfaces as well as stochastic loads, such as electric vehicles, this robustness can no longer be counted on. Thus operators are more interested in designing the wide-area controller $\bm{u}$ by online learning.
 
 One choice is to design a LQR controller for $\bm{u}$, as shown in \cite{5}. For this, we will assume the state $\bm{x}(t)$ to be available for feedback. This can be done by placing PMUs at geometrically observable set of buses, so that the voltage and current phasors at every generator bus are computable. A decentralized unscented Kalman filter is assumed to be installed at every generator. The computed (or measured if a PMU is already at the generator bus) values of the bus voltage and current phasors are used by the Kalman filter to estimate the generator state vector $\hat{\bm{x}}_i$. Assuming that the KF runs continuously and is sufficiently faster than the generator dynamics, for the rest of the paper we will simply assume $\hat{\bm{x}}_i=\bm{x}_i$. For more details on this KF please see \cite{pal}.
\subsection{Optimal Wide-Area Control}
The wide-area damping control problem for the power system model (\ref{eq8}) is posed as a LQR problem, i.e. find matrix $\bm{K} \in \mathbb{R}^{m \times n}$ such that the state-feedback control $\bm{u}(t)=-\bm{Kx}$ minimizes the quadratic energy function:
\begin{equation} \label{eq9}
J = \int_{0}^{\infty}(\bm{x}^{T}(t)\bm{Q}\bm{x}(t) + \bm{u}^{T}(t)\bm{R}\bm{u}(t))dt 
\end{equation}
where $\bm{Q} \geq 0$ and $\bm{R} > 0$ are design matrices with appropriate dimensions. In the vector form, the controller can be expressed as
\begin{equation} \label{eq10}
\begin{bmatrix}
u_{1}(t)\\
u_{2}(t)\\
\vdots\\
u_{n}(t)\\
\end{bmatrix} = 
\begin{bmatrix}
\bm{K}_{11} & \bm{K}_{12} & \cdots & \bm{K}_{1n}\\
\bm{K}_{21} & \bm{K}_{22} & \cdots & \bm{K}_{2n}\\
\vdots\\
\bm{K}_{n1} & \bm{K}_{n2} & \cdots & \bm{K}_{nn}\\
\end{bmatrix} 
\begin{bmatrix}
\bm{x}_{1}(t)\\
\bm{x}_{2}(t)\\
\vdots\\
\bm{x}_{n}(t)\\
\end{bmatrix} 
\end{equation}
where the sub-matrix $\bm{K}_{ij}$ indicates feedback gain from the states of generator $j$ to the controller of generator $i$, while $\bm{K}_{ii}$ is the self-feedback gain for generator $i$. Following Hamiltonian theory, the controller $\bm{K}$ can be designed by solving the algebraic Riccati equation \cite{OP1}.
When $\bm{K}$ is optimal, every $\bm{K}_{ij}$ and $\bm{K}_{ii}$, in general, are non-zero matrices.
We define feedback links from states to control input within one generator as self-links and the feedback links between different generators as communication links. Solution of LQR will usually result in a dense $ \bm{K} $ requiring a communication link from every state to every control input. Such dense communication graph requires large volume of data exchange among the generators. In order to reduce the communication cost \cite{jsac}, we therefore impose an extra constraint of reducing the cardinality of $\bm{K}$ to make it sparse and reduce the number of communication links
\begin{equation} \label{eq11}
s = \mbox{Card}_{\mbox{off}}(\bm{K}) = \sum\nolimits^{n}_{i,j=1, i \neq j}\mbox{nnz}(\bm{K}_{ij})
\end{equation}
where nnz(.) operator returns the number of nonzero elements of a matrix. The self-feedback gains $\bm{K}_{ii}$ are not counted in the definition (\ref{eq11}) due to their negligible cost.

\subsection{Model Uncertainties}
We assume that the exact values of the matrices $\bm{A}$, $\bm{B}$ in (\ref{eq8}) are not known to the power system operator. Only a nominal pair ($\bm{A_{0}, B_{0}}$) is known.  Typical uncertainties in $\bm{A}$, $\bm{B}$ in an actual power system may result from various unknown parameters such as inertias of the generators, especially when power electronic converters are added resulting in low-inertia equivalents \cite{emi}, or exact values of the line reactances, especially when series compensation may be used in long transmission lines for certain unforeseen contingencies, or even the internal time constants of the generator circuits,  uncertainties in load dynamics and associated load control mechanisms, unpredictable plug-and-play dynamics of converters, intermittent generation profiles of hundreds of wind, solar and storage devices, and so on.

We refer to the LQR controller based on the nominal model $\bm{A_{0}, B_{0}}$ as the \textit{mismatched } LQR $\bm{K}_{mis}$. When the actual $\bm{A, B}$ deviate significantly from the nominal model matrices $\bm{A_{0}, B_{0}}$, the performance of the mismatched LQR controller suffers and can even become unstable as illustrated in section 4. Thus, we investigate RL under constrained communication cost given the knowledge of the nominal model $\bm{A_{0}, B_{0}}$, but not of the actual model $\bm{A}$, $\bm{B}$.

\section{RL control for WAC}
Reinforcement learning has been proposed as a tool to implement optimal control for unknown or uncertain systems \cite{RL4}. A combination of Q-learning and adaptive dynamic programming is proposed in \cite{V17}, which provides an actor-critic structure capable of learning the optimal control policy for completely unknown, continuous-time dynamic systems using value iteration. This algorithm is implemented online using  state and control input measurements of the system. Unlike a general RL problem, where $\bm{K}$ is learnt online starting from any arbitrary initial guess, uncertainties in power systems are typically not that unstructred and drastic. This means that if ($\bm{A_{0}, B_{0}}$)  is the model during one contingency, and ($\bm{A_{1}, B_{1}}$) is the model during a contingency that occurs within a few hours, then most probably only a few entries of ($\bm{A_{1}, B_{1}}$) will be different from those of ($\bm{A_{0}, B_{0}}$) as only a few line and generator parameters may have changed between the two events. The initial guess for $\bm{K}$  can therefore be picked as the controller from the previous contingency. If the difference between the models is indeed not significant, then this choice would expedite the convergence of this loop considerably. As the discrepancy between the two models grows, choosing $\bm{K}$ corresponding to the nominal  model ($\bm{A_{0}, B_{0}}$) still increases the convergence speed of RL significantly relative to a random guess as will be illustrated in section 4. The notation used in this section is summarized in Table 1.

\begin{table}[!htb]
\caption{Notation used in Algorithm 1.}
\footnotesize{
\begin{tabular}{L{2.7cm} L{5.3cm}}
\toprule
\textbf{Term} & \textbf{Definition} \\ \midrule

$\alpha_{c}$ & Critic convergence speed coefficient.\\ \midrule

$\alpha_{a}$ & Actor convergence speed coefficient.\\ \midrule

$ \bm{U} = [\bm{x}^{T} \bm{u}^{T}]^{T}$ & Concatenated vector of states and control inputs.\\ \midrule

$\bm{\Phi}(t) = \bm{U}(t) \otimes \bm{U}(t)$ & Quadratic basis vector of states and control inputs. \\ \midrule

$\bm{\sigma} = \bm{\Phi}(t) - \bm{\Phi}(t-T)$ & Change in the basis vector after time-step T. \\ \midrule

$ \bm{G} = \begin{bmatrix}
   \bm{G}_{11}   &  \bm{G}_{12} \\
   \bm{G}_{21}   &  \bm{G}_{22} \\
\end{bmatrix} $ & Kernel $\bm{G} \in \mathbb{R}^{(n+m) \times (n+m)}$ \\ \midrule
$\bm{W}  = vech(Qf) $ & Critic vector \\ \midrule

$\bm{K}_{SP} $ & Actor vector\\ \midrule

$e_{c} = \bm{\hat{W}}^{T} \bm{\sigma} + \quad  \int_{t-T}^{t} (\bm{x}^{T}\bm{Qx} + \bm{u}^{T}\bm{Ru})d\tau$ & Critic error\\ \midrule

$e_{a} = \hat{\bm{K}_{SP}}^{T} \bm{x} + \hat{\bm{G}}_{22}^{-1}\hat{\bm{G}}_{21}\bm{x}$ & Actor error\\ \midrule

$\bm{\dot{W}} = -\alpha_{c} \frac{\sigma}{(1+\sigma^{T} \sigma)^{2}} {e_{c}}^{T}$ & Critic update\\ \midrule

$\bm{\dot{K}}_{SP} =  -\alpha_{a}\bm{x}{e_{a}}^{T}$ & Actor update\\ \midrule

$ Qf(x,u) = \frac{1}{2} \bm{U}^{T} \bm{G} \bm{U} $ & Q-function \\ \midrule

$\bm{u}(t) = arg\min_{u} Qf(\bm{x},\bm{u}) = -\bm{G}_{22}^{-1}\bm{G}_{21}x $ & Optimized control input \\ \midrule

$ vech(\bm{.}) $ & Half vectorization operator, stacks elements of the upper triangular part of a matrix into a vector, multiplying diagonal elements by 2. \\ \midrule

$\Vert \bm{K} \Vert_{2}$ & Frobenius norm of the matrix $\bm{K}$, defined by trace($\bm{K}^{T}\bm{K}$).\\ \midrule

supp($\bm{K}$) & The support set of the matrix $\bm{K}$, i.e., the set of indices of the nonzero entries of matrix $\bm{K}$.\\  \midrule

$[\bm{K}]_{s}$ & The  matrix  obtained  by  preserving  only  the $s$ largest-magnitude  entries  of  the  matrix $\bm{K}$,  and setting all other entries to zero.\\  \midrule

$\nabla_{\bm{K}} (\lVert e_{a}\lVert^{2})$ & The gradient of $(\lVert e_{a}\lVert^{2})$ w.r.t $\bm{K}$. Assuming $\bm{K} \in \mathbb{R}^{m \times n}$, $\nabla_{\bm{K}} (\lVert e_{a}\lVert^{2})$ is given by $m \times n$ matrix with the elements $[\nabla_{\bm{K}} (\lVert e_{a}\lVert^{2})]_{ij} = \partial \lVert e_{a}\lVert^{2}/ \partial \bm{K}_{ij}$. \\  \midrule

$\Delta_{nwt}(\bm{K},\tau)$ & The restricted Newton step of an arbitrary function $f(\bm{K})$ at matrix $\bm{K} \in \mathbb{R}^{m \times n}$ under structural constraint supp($\bm{K}$) $\subset \tau$. First, all elements of $\bm{K}$ is stacked in vector $\bm{x}$, and the function $g(\bm{x})$ is defined as $g(\bm{x}) \triangleq f(\bm{K}) $. Then  the $mn \times 1 $ restricted  Newton step vector $\Delta_{nwt}(\bm{x},\tau)$ of $g(\bm{x})$ at $\bm{x}$ is computed using the conjugate gradient method \cite{lin2013design}. The vector  $\Delta_{nwt}(\bm{x},\tau)$ is then converted back to $m \times n$ matrix by  stacking consecutive $m \times 1$ segments of $\Delta_{nwt}(\bm{x},\tau)$.\\ \bottomrule

\end{tabular}}
\vspace{-0.5cm}
\end{table}

\subsection{Sparsity-constrained WAC using RL}

First, we briefly review the RL algorithm in \cite{V17}. The two-step learning iterative process starts from randomly generated actor and critic vectors and iterates until these vectors converge to their optimal values. The critic approximator is responsible for estimation of the Q-function while the actor approximator aims to find the optimal control policy. In each step, the actor selects a control policy ($\bm{K}_{RL}$) and applies $\bm{u} = -\bm{K}_{RL} \bm{x}$ to the unknown system (\ref{eq8}) in real-time. The critic evaluates the performance of the control policy applied by the actor using state and control input measurements. This evaluation is then used by the actor to update its control policy. This process  continues until the actor update results in unchanged $\bm{K}_{RL}$ within a desired amount of error. 

The critic and actor update rules in \cite{V17} are implemented by gradient descent and do not require the knowledge of the system model. Thus the algorithm in \cite{V17} converges to the optimal LQR controller for a completely unknown system. However, this method has long convergence time. Since in power systems partial system knowledge is available to the designer, we employ the known matrices $\bm{A_{0}, B_{0}}$ of the nominal model to initialize the RL algorithm, which can reduce the convergence time. Note that using ($\bm{A_{0}, B_{0}}$), however, does not classify Algorithm 1  as a model-dependent algorithm. The algorithm is still model-free; it is only \textit{assisted} by the knowledge of the nominal model so that one can expedite the learning phase.
In addition, to add sparsity in $\bm{K}$, we propose the control design problem as follows:

\begin{align} \label{eq12}
\min_{\bm{K}} \quad & J(\bm{K}) \\
\mbox{s.t.} \quad &  \mbox{Card}_{\mbox{off}}(\bm{K}) \leq s \nonumber \\
\bm{\dot{x}}(t) = & \mbox{ } \bm{Ax}(t) + \bm{Bu}(t),\quad \bm{x}(0) = \bm{x}_{0} \nonumber \\
\bm{u}(t) = &  -\bm{Kx}(t) \nonumber
\end{align}
\noindent where $\bm{A}$ and $\bm{B}$ are unknown matrices. The constrained optimization (\ref{eq12}) results in a sparse $\bm{K}$ matrix denoted by $\bm{K}_{SP}$, which has at most $s$ communication links. We employ GraSP algorithm \cite{grasp} in the actor update step of\cite{V17} instead of gradient descent to impose this constraint in RL. GraSP was shown to find sparse solutions for a wide class of optimization problems. The details of the RL controller for WAC given  the sparsity constraint are provided in Alg. 1.

The disturbance happens in step 1, followed by all nodes sending their state information to the central controller in step 2. In step 3, the initial critic and actor vectors are generated. First, we solve the Riccati equation for the nominal model $\bm{A_{0}, B_{0}}$ to find positive definite $\bm{P}_{0}$ 
\begin{equation} \label{eq13}
\bm{A}_{0}^{T}\bm{P} + \bm{P}\bm{A}_{0} - \bm{P}\bm{B}_{0}\bm{R}^{-1}\bm{B}_{0}^{T}\bm{P} + \bm{Q} = \bm{0} 
\end{equation}
and then we form the kernel $\bm{G}$
\begin{equation} \label{eq14}
\bm{G} = \begin{bmatrix}
   \bm{P}_{0}\bm{A}_{0} + \bm{A}_{0}^{T}\bm{P}_{0} + \bm{Q} + \bm{P}_{0}   &  \bm{B}_{0}\bm{P}_{0} \\
    \bm{P}_{0}^{T}\bm{B}_{0}       & \bm{R} \\
\end{bmatrix} 
\end{equation}
Next, in step 3 of Alg. 1, using $\bm{U}_{0}^{T} = [\bm{x}_{0}^{T} \mbox{ } \bm{u}_{0}^{T}]$, we find the Q-function and  $\bm{W}_{0}$
\begin{equation} \label{eq15}
\bm{W}_{0} = vech(Qf(\bm{x}_{0}, \bm{u}_{0}))
\end{equation}
The initial actor matrix $\bm{K}_{SP}^{0}$ is given by \cite{V17}
\begin{equation} \label{eq16}
\bm{K}_{SP}^{0} = \bm{R}^{-1}\bm{B}_{0}^{T}\bm{P}_{0}
\end{equation}

\begin{algorithm}[!htb]
\caption{ RL for WAC  under sparsity constraint $s$}

\begin{algorithmic}[1]
\State Time of disturbance: $t = 0$, iteration $i=0$.
\State All nodes send states to central controller.
\State Central controller receives $\bm{x}_{0}  $ and forms initial critic $\bm{W}^{0}$ (\ref{eq15}) and actor $\bm{K}_{SP}^{0}$  (\ref{eq16}).
\While{${\Vert \bm{K}_{SP}^{i+1} - \bm{K}_{SP}^{i}\Vert }_{2}\ge \epsilon_{K}$}
\State Calculate control input $\bm{u}(t)$ (\ref{eq10}) 
\If{ $t \leq T_{PE}$ } \Comment{ add exploration noise}

\State $ \bm{u}(t) \longleftarrow  \bm{u}(t) + \bm{u}_{PE}(t)$

\EndIf
\State Apply $\bm{u}(t)$ to the system $\bm{\dot{x}} = \bm{Ax} + \bm{Bu}$ for $T$ sec.
\State All nodes send state and input measurements to central controller.
\If{${\Vert \bm{W}^{i+1} - \bm{W}^{i}\Vert }_{2}\ge \epsilon_{W}$} 
\State Update critic vector ($\bm{W}$) (Table 1)

\EndIf

\State calculate the gradient of $e_{a}$ $\rightarrow g = \nabla_{\bm{K}_{SP}^{i}} (\lVert e_{a}\lVert^{2})$

\State Keep $2s$ largest entries of g $\rightarrow Z = \mbox{supp}([g]_{2s})$

\State Merge supp($\bm{K}_{SP}^{i}$) and Z $\rightarrow \tau = Z + \mbox{supp}(\bm{K}_{SP}^{i})$

\State Descend based on restricted Newton step $ \rightarrow \dot{\bm{K}_{SP}^{i+1}} = \lambda \Delta_{nwt}(\bm{K}_{SP}^{i},\tau)$

\State Prune extra entries: keep $s$ largest elements of $\bm{K}_{SP}^{i+1}$ and set others equal to zero $\rightarrow \bm{K}_{SP}^{i+1} = [\bm{K}_{SP}^{i+1}]_{s}$


\State $i \longleftarrow i+1$
\EndWhile
\State \textbf{return} $\bm{K}_{SP}^{i+1}$\Comment{$\bm{K}_{SP}^{i+1}$ is sparse with $s$ non-zero elements}
\end{algorithmic}
\end{algorithm}
The actor update loop begins in step 4 and extends until end of the Algorithm 1, where $\epsilon_{K}$ is the desired actor error. The control input is calculated in step 5. Steps 6 and 7 implement addition of the exploration noise, where $T_{PE}$ indicates the duration of the exploration noise $\bm{u}_{PE}$, chosen to provide sufficient persistence of excitation for the convergence of the critic approximator. The noise signal $\bm{u}_{PE}(t)$ is given by the sum of sinusoids with a sufficient number of frequencies \cite{V17}. In step 8, the control input $\bm{u}(t)$ is applied to the system, and the resulting state and input measurements are sent to the central controller in step 9. Steps 10 and 11 update the critic as depicted in Table 1. Since the critic must converge faster than the actor, we choose the learning rate of the critic to be much greater than that of the actor $(\alpha_{c} \gg \alpha_{a})$. Moreover, $\epsilon_{W}$ is the desired error for the critic \cite{V17}. Steps 12-17 of Alg. 1 implement GraSP on the update of actor, which limits the directions in which this update is performed \cite{grasp}. In each iteration, $\bm{K}_{SP}$ is extended along its $2s$ steepest gradient-descent directions (step 13). Then the set of descent directions is created by merging the indices of non-zero elements of $\bm{K}_{SP}$ found in the previous iteration and $2s$ largest elements of the gradient. As a result, the direction of descent will have at most $3s$ elements (step 14). After descent based on these directions, an $\ell_{0}$-norm is applied to the resulting $\bm{K}_{SP}$ to remove the extra entries and ensure $\mbox{Card}_{\mbox{off}}(\bm{K}_{SP}) = s$ (step 16).
\subsection{Timeline and Cyber-Physical Implementation}
\begin{center}
\begin{figure}[!tp]
      \centering
      \includegraphics[width = 3.4in]{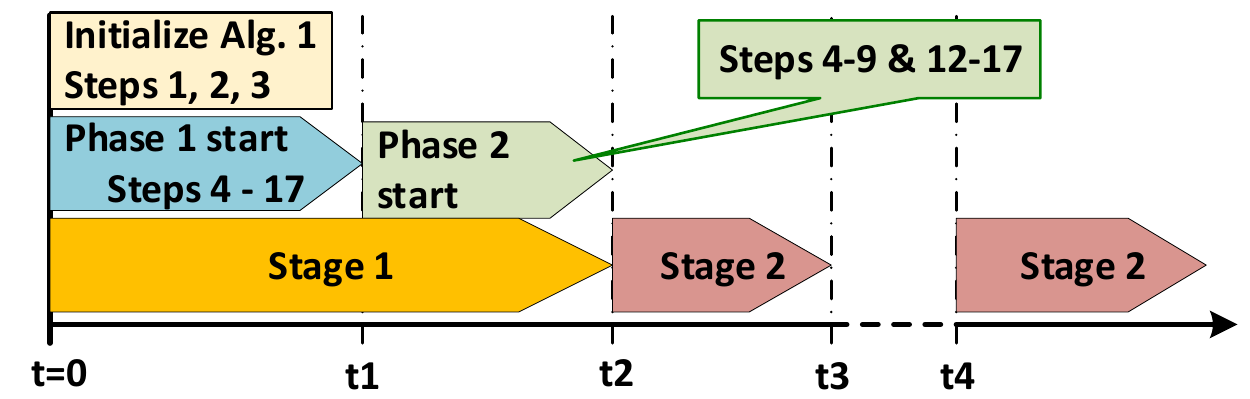}
      \caption{Timeline for Alg. 1.}
      \label{time}
        \vspace{-2em}
 \end{figure}
\end{center}

\begin{center}

\begin{figure*}[!tp]
 \hspace*{-.5cm}
\vspace*{-0.5cm}
      \centering
      \includegraphics[width = 6.8in]{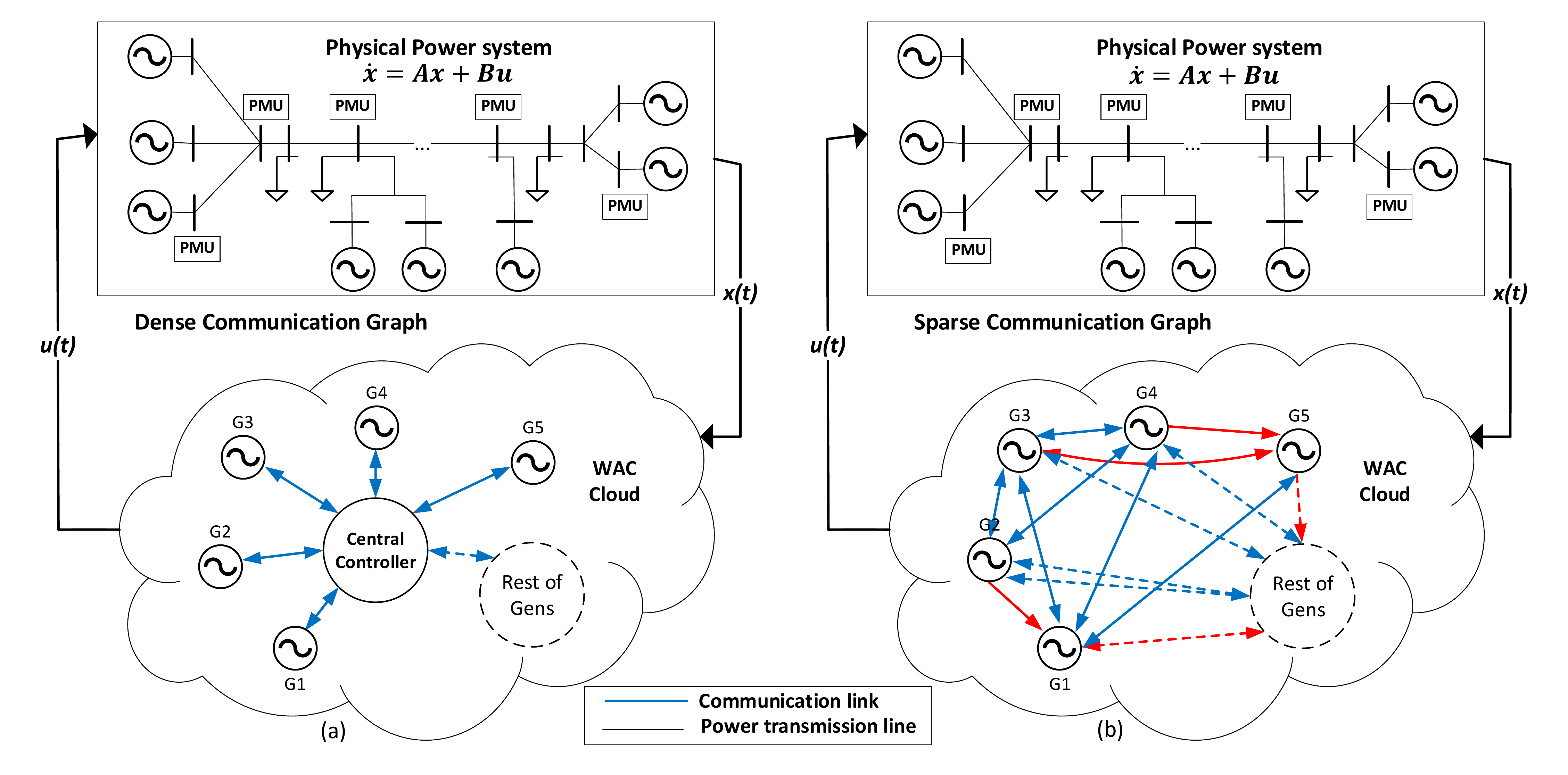}
      \caption{Stages of the learning algorithm and implementation of the sparse wide-area controller; $ t_2$ is the convergence time of stage 1. (a) Learning of $\bm{K}_{SP}$, $t = 0$ to $t_{2}$ (Stage 1). (b) Implementation of the sparse wide-area controller $\bm{K}_{SP} $ (Stage 2). The blue and red lines indicate the required links for dense communication while only the red lined are required for sparse implementation of WAC. The dashed lines indicate communication links between indicated generators and remaining generators which are illustrated by a dashed circle. Hence, the dashed line may actually be more than one link.}
 \label{cps}    
\end{figure*}
\end{center}

Fig. \ref{time} shows the Timeline, while Fig. \ref{cps} shows the closed-loop configuration as a cyber-physical system (CPS) and the stages of design and implementation of the proposed algorithm. As shown in Fig. \ref{time}, in stage 1, the learning Algorithm 1 finds a sparse controller ($\bm{K}_{SP}$) suitable for damping oscillations of the actual system. In stage 2, we apply $\bm{K}_{SP}$ right after learning it to damp oscillations. We also apply it at $t_{4}$, when new disturbances occur, assuming the system model remains unchanged. This assumption is based on typical power system conditions for disturbances within a short time interval of each other. In stage 1, centralized computation is carried out to learn the controller as shown in Fig. 2(a). Stage 1 consists of two phases. In \textit{phase 1} (steps 4 to 18 of Alg. 1), both the critic and the actor estimators are updating. While the actor is sparse with the sparsity constraint $s$, its sparsity pattern can change, i.e. the indices of non-zero elements in $\bm{K}_{SP}$ might change. When the critic converges to its final value (sooner than the actor), the \textit{phase 1} finishes, and the structure of the sparsity-constrained actor is fixed. 

In \textit{phase 2}, only the actor parameters are updated while the sparsity pattern, which determines the communication graph between the nodes of the plant, is fixed. Phase 2 extends over steps 4-18, but excludes steps 10-11, i.e. the critic vector update. Finally, Fig. 2(b) illustrates stage 2. It shows a sparse controller produced by the learning algorithm for a typical multi-machine grid. In this case, the sparse feedback gain matrix $\bm{K}_{SP}$ is fixed and known to each generator. Thus there is no need for the central controller, and each generator can compute its own control input $u_{i}$ using state measurements from other generators and $\bm{K}_{SP}$, thus implementing eq. (\ref{eq10}) in a distributed fashion. 
\begin{center}
\begin{figure*}[!tp] \label{figure3}
\vspace*{-1.5cm}
      \includegraphics[width = 7 in]{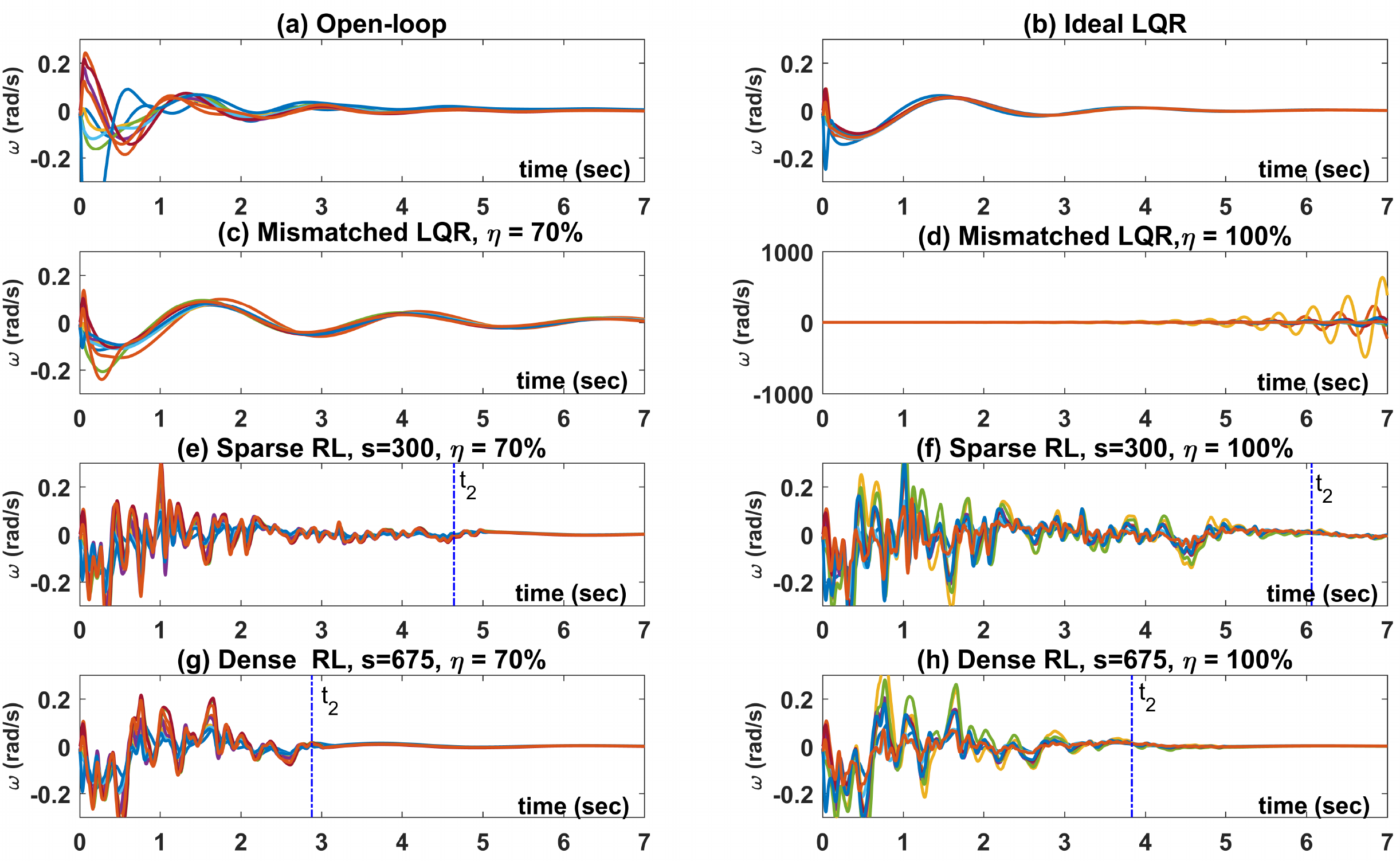}
      \caption{Performance of different controllers in damping oscillations; $\eta$ indicates the uncertainty level, and $t_2$ is the end of stage 1 (learning).}
      \label{figure4-3}
 \end{figure*}
 \end{center}{}
\vspace*{-2cm}
\section{Numerical Results}
The IEEE 39-bus model of New England power system is used in this section to study the effectiveness and performance of the proposed algorithm. This model consists of 10 synchronous generators. The state-space model in the form of (\ref{eq10}) has the state vector of size 75, i.e. $\bm{A} \in \mathbb{R}^{75 \times 75}$ and 9 control inputs, resulting in $\bm{B} \in \mathbb{R}^{75 \times 9}$ and $\bm{K} \in \mathbb{R}^{9 \times 75}$. The details of the model can be found in \cite{5}. The feedback gain matrix has 675 entries. Following the discussion in section 2.C, the uncertainty of the power system model is limited to the parameters corresponding to inertias of the machines and line reactances, which results in 1132 uncertain parameters in $\bm{A}$.  Moreover, $\bm{B}$ matrix has 9 non-zero entries with known indices and uncertain values. The level of uncertainty in both $\bm{A}$ and $\bm{B}$ is measured by the parameter $\eta$. To generate uncertain parameters, we randomly vary the corresponding entries of $\bm{A, B}$ using the uniform distribution within $\eta \%$ of their original values in $\bm{A}_{0}$ and $\bm{B}_{0}$ matrices, respectively. Finally, the duration of the exploration noise $T_{PE}$ in step 6 of Algorithm 1 is 2 seconds, and the error terms $\epsilon_{K}$ and $\epsilon_{W}$ in steps 4 and 10 are equal to $10^{-5}$. 

Several WAC scenarios are simulated using two different $\eta$ values of 70\% and 100\%. We compare four controllers: the \textit{ideal} LQR controller ($\bm{K}_{lqr}$), designed assuming the knowledge of the actual system model $\bm{A, B}$; the \textit{mismatched} LQR controller ($\bm{K}_{mis}$), designed for the nominal system $\bm{A}_{0}$, $\bm{B}_{0}$ and applied to the actual system $\bm{A, B}$; the \textit{dense learning} controller ($\bm{K}_{dense}$) found by Algorithm 1 for the maximum value of $s =  675$; the \textit{sparse learning} controller ($\bm{K}_{SP}$) designed using Algorithm 1 for a range of $s$ values ($ 100 \leq s \leq 675$). Moreover, performance of the open-loop system controlled by local PSS (power system stabilizers), designed for $\bm{A, B}$, is shown for comparison. Note that although by definition self-links does not incur any communication cost, the algorithm  is  capable of removing them if $s$ is sufficiently small.

As described earlier, we assume that only the nominal model  $\bm{A}_{0}$, $\bm{B}_{0}$ is known to the designer. Modeling the fault as an impulse input that is cleared at $t=0$, we start the simulation from the initial state $\bm{x}(0)$ at $t=0$. Figure 3 shows the time response of the generator speeds for the different controllers and uncertainty levels. The convergence time of the learning algorithm (end of stage 1) is denoted as $t_{2}$ in Fig. 3 (e-h). Note that when $\eta = 100\%$, the \textit{mismatched} LQR designed for the nominal model becomes unstable while the RL controller suppresses oscillations in both sparse and dense cases. The convergence time of Algorithm 1 increases with the level of uncertainty for both sparse and dense controllers. 

Next, we apply $\bm{K}_{SP}$ learned in stage 1 (fig \ref{cps}(a)) to damp the oscillations caused by new disturbances (see fig \ref{time}) for both uncertainty levels. Figure 4 indicates the rotor speed of the generators in this scenario. Note that the mismatched LQR is still unstable when $\eta = 100\%$ while $\bm{K}_{SP}$ successfully damps the oscillations for both $s$-values. We found that $\bm{K}_{SP}$ does not depend on the initial condition $\bm{x}_{0}$ at $t=0$. Therefore, $\bm{K}_{SP}$ designed by Alg. 1 performs well for any incoming disturbance. 

The performance of the controllers in terms of the increase in the closed-loop energy $J$ in  (\ref{eq9}) compared to the \textit{ideal} LQR is reported in Table 2. The $J$-value is calculated from $t=0$ (when the initial disturbance happens) till $t=10$ seconds when the oscillations are practically damped. Note that in this case the overall energy is not significantly affected by learning due to relatively short duration of Stage 1 when the nominal knowledge of the system is used. It can be seen that the values of $J$ in figures 3 and 4 are comparable for fixed values of $\eta$. 

When Algorithm1 does not employ the sparsity constraint and the nominal model, it is identical to the RL method in \cite{V17}, which was proven to converge to the optimal stable solution of LQR (Theorem 2, \cite{V17}). Addition of the nominal model does not change the convergence result since the latter applies to any initial control matrix including (\ref{eq16}). However, sparse control problems elude theoretical convergence guarantees \cite{5, jsac}. If an ideal closed-form solution to (\ref{eq12}) was available for known system matrices (similar to the Riccati equations (\ref{eq13}) for the LQR problem), the arguments of \cite{V17} would apply to prove convergence of the sparsity-constrained RL design to this ideal solution.

In practice, even when the system matrices $\bm{A}, \bm{B}$ are known, convergence of GraSP is not assured for a given value of $s$ \cite{jsac}, implying that Algorithm 1 of this paper does not necessarily converge. However, if it converges, step (15) guarantees that $\bm{K}_{SP}$ satisfies the following property
\begin{equation}
\nabla_{\bm{K}} (\lVert e_{a}\lVert^{2})|_{\mbox{supp}(\hat{\bm{K}}_{SP})}({\bm{K}}_{SP} =  \hat{\bm{K}}_{SP}) = 0,
\end{equation}
which is the weak necessary condition for the optimality of (\ref{eq12}) \cite{beck2013sparsity}. Thus, the feedback matrix $\bm{K}_{SP}$ at convergence of Algorithm 1 corresponds to the global or a local minimum of the optimization problem (\ref{eq12}). 
Extensive numerical studies show that the proposed algorithm converges in many practical scenarios for all uncertainly values and sparsity ranges $s = 100-675$, and the closed-loop system is observed to be stable.

As expected, the value of the objective function (\ref{eq9}) increases as the sparsity level grows (decreasing $s$). This is due not only to sub-optimality of the sparse controller, but also to increased duration of learning,  and thus more persistent oscillations for lower $s$ values.
\begin{center}
\begin{figure}[!htb] 
  \vspace{-1.5em}
      \centering
      \includegraphics[width = 8.6 cm]{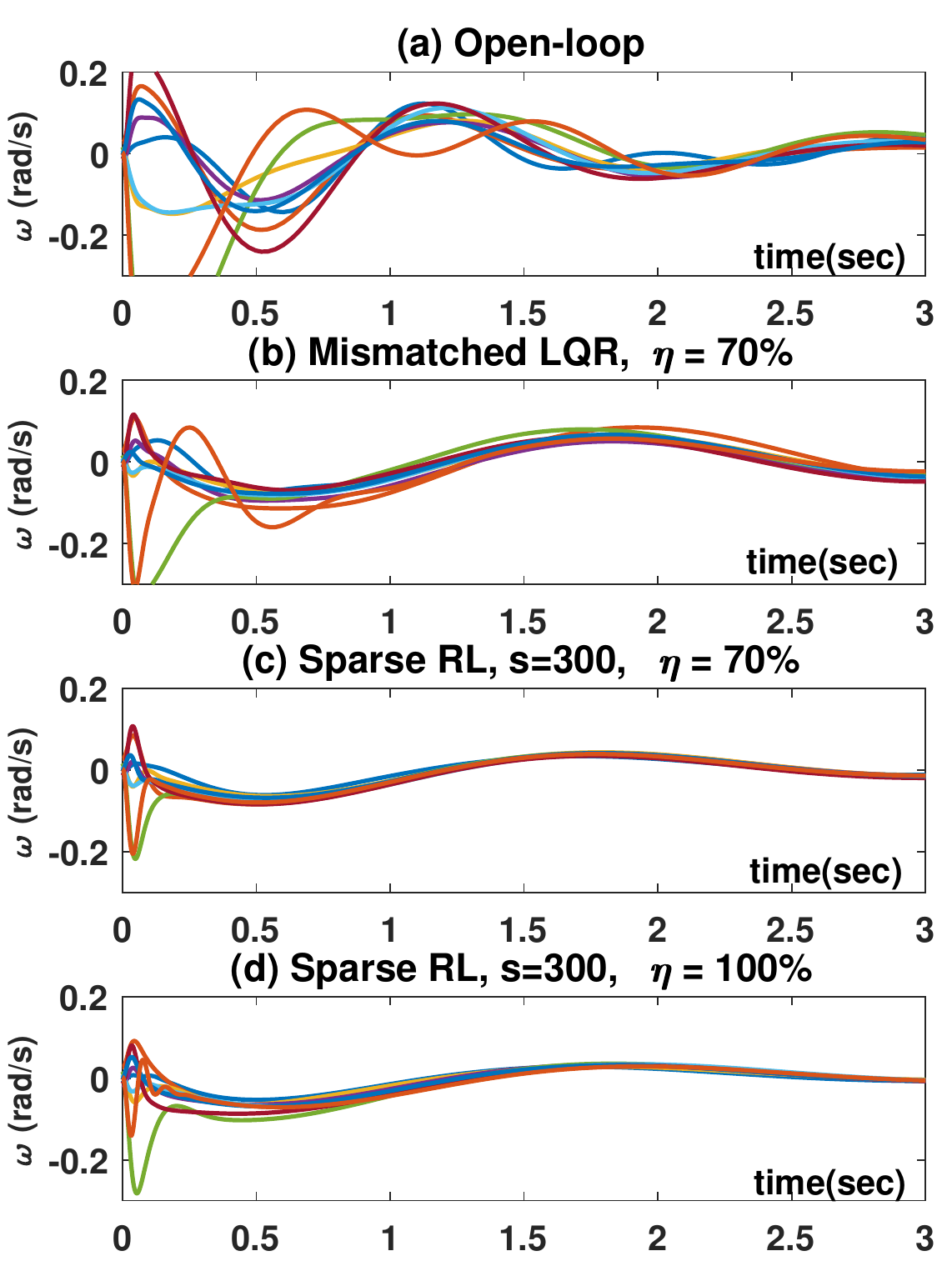}
      \caption{Performance for future disturbances using previously learnt controller; $\eta$ indicates the uncertainty level .Note that the mismatched LQR becomes unstable for $\eta= 100\% $ (not shown).}
      \label{figure4-4}
      \vspace{-1.5em}
 \end{figure}
\end{center}
\begin{center}
\begin{table}[!htb]\small
\vspace{-2 em}
\caption{Increase in $J$-values compared to ideal LQR.}
\begin{tabular}{ |c|c|c|c|c| } 
\hline
Figure & 3 & 3 & 4 & 4 \\
\hline
$\eta$ & 70\% & 100\% & 70\% & 100\% \\
\hline
{\Centerstack{Open-loop \\ $J$}} & 1470\% & 1470\% & 221.49\% & 221.49\% \\
\hline
{\Centerstack{Mismatched \\ LQR $J$}} & 13.68\% & $\infty$ & 16.55\% & $\infty$ \\
\hline
{\Centerstack{Sparse RL $J$\\ $s=300$}} & 6.84\% & 11.43\% & 5.14\% & 6.87\% \\
\hline
{\Centerstack{Dense RL $J$\\ $s=675$}} & 1.8\% & 3.27\% & 0.23\% & 0.23\% \\
\hline
\end{tabular}
\end{table}
\end{center}
\vspace{-1 em}
Finally, the convergence time of Alg. 1 depends on two factors. First, it increases with the uncertainty level $\eta.$ As the deviation between the nominal model and the actual model increases, the convergence time also increases sharply since the initial critic and actor do not approximate their optimal values closely for large $\eta$-values. Moreover, decreasing the sparsity constraint $s$ results in slower convergence. For $s < 100$, the convergence time increases dramatically, thus precluding utilization of extremely sparse controllers in WAC when employing RL. If Alg. 1 was initialized with randomly generated actor and critic as in \cite{V17}, the convergence time would be on the order of hours and thus exceed significantly the acceptable range for WAC applications. Moreover, the exploration noise duration would increase greatly as well, adding extra oscillations to the system and degrading the quality of service during that period. Hence, using the knowledge of the nominal model to initialize the RL algorithm enables its application in WAC.
\section{Conclusions}
We proposed a sparse wide-area control design for power systems using online, data-driven reinforcement learning. First, the convergence time was reduced significantly by using the knowledge of the nominal model. Second, the communication cost of WAC was reduced by sparsifying the controller using the GraSP method. The effectiveness of the proposed controller was illustrated on the IEEE New England power system model with uncertain parameters. It was demonstrated that the proposed sparse controller successfully damps the wide-area oscillations even for highly uncertain models, while the  LQR controller matched to the nominal model destabilizes the system as the uncertainty level grows. Future work will further address the convergence properties of the proposed RL algorithm and extend  this centralized learning method to distributed and multi-agent implementations.











\bibliographystyle{IEEEtran}
\bibliography{ref}

\end{document}